\documentstyle[12pt]{article}

\newcommand{\be}{\begin{equation}}
\newcommand{\ee}{\end{equation}}
\newcommand{\bq}{\begin{eqnarray}}
\newcommand{\eq}{\end{eqnarray}}
\newcommand{\nl}{\newline}

\begin{document}

\begin{flushright}
QMW-PH/96-7
\end{flushright}

\begin{center}
{\Large \bf N=1 M-theory-Heterotic Duality in Three Dimensions and
Joyce Manifolds.}
\newline

{\Large B.S. Acharya\footnote{e-mail: r.acharya@qmw.ac.uk. Work supported
by PPARC.}}
{\it Queen Mary and Westfield College, Mile End Road, London. E1 4NS.}
\end{center}

\begin{abstract}
It is argued that $M$-theory compactifications on {\it any}
of Joyce's $Spin(7)$ holonomy 8-manifolds are
dual to compactifications of 
heterotic string theory on Joyce 7-manifolds
of $G_2$ holonomy.
\end{abstract}

\newpage
\section{Introduction.}

Over the past year or so, our perception of string theory has dramatically
altered \cite{cmh,wit}. The emergence, of the new dimension which has
opened up in our understanding, can perhaps be attributed to 
the magical $M$-theory
\cite{M,hor}. One virtue of this viewpoint is that one can ``understand''
connections between string theories \cite{cmh,wit,kach,vaf,har,klemm,quev}
from an eleven dimensional perspective \cite{M,hor}.

In \cite{Ach}, we
presented an ansatz to construct dual $M$-theory/heterotic compactifications, 
starting
from $M$-theory/heterotic duality in seven dimensions. In \cite{Ach} we
focussed on dual compactifications with $N=1$ supersymmetry in four 
dimensions. This paper is devoted to constructing $M$-theory/heterotic duals
with $N=1$ supersymmetry in three dimensions, by applying the same ansatz as 
in \cite{Ach}. In fact, we will construct heterotic duals (on Joyce
7-manifolds of $G_2$ holonomy) for $M$-theory compactifications on all
known Joyce 8-manifolds of $Spin(7)$ holonomy. 

It turns out, following
the recent work of Sen \cite{sen}, that one can {\it explain} 
the ansatz presented in \cite{Ach}
in the following way. Consider $M$-theory on an eight-torus, $T^8$, with
coordinate labels $x_1,x_2,...x_8$. This is equivalent \cite{wit} to
Type IIa string theory on $T^7$, where the coordinate labels of $T^7$
are any seven element subset of $x_1,x_2,...x_8$. For definiteness, we
choose $T^7$ to be labelled by $x_1,x_2,x_3,x_5,x_6,x_7,x_8$. 
Take a $Z_2$ orbifold of the $M$-theory compactification,
where the $Z_2$ acts on $x_1,x_2,x_3,x_4$ as reflection. This gives
$M$-theory on an orbifold limit of $K3{\times}T^4$. In \cite{sen}, it was
argued that this orbifold commutes with Type IIa/$M$-theory duality and
is equivalent to TypeIIa theory on ${T^7}/{{Z_2}^{\prime}}$, 
where ${Z_2}^{\prime}$ acts as reflection on $x_1,x_2,x_3$ combined with
world-sheet parity and the element $(-1)^{F_L}$, where $F_L$ is the
left moving fermion number operator in the IIa theory. One can then
make a $T$-duality transformation on the three circles labelled by $x_1,x_2
,x_3$, which inverts their radii. This gives the Type IIb theory on
a ${T^7}/{\Omega}$ orbifold, where $\Omega$ is the world-sheet parity 
operator. This is by definition Type I theory on a $T^7$ which shares the
same coordinate labels as its equivalent TypeIIa compactification above.
One can then use Type I-heterotic duality \cite{het} to map this
to the heterotic string compactified on $T^7$. The crucial point, for
what follows, is that the seven coordinate labels of the heterotic string
on $T^7$ are a subset of the eight coordinate labels of the $M$-theory
compactification on ${{T^4}/{Z_2}}{\times}T^4$. 

The analysis presented above is an explanation of the first part of the
ansatz presented in \cite{Ach}. As in \cite{Ach}, we wish to consider taking
further $Z_2$ orbifolds of the three dimensional $N=8$ $M$-theory
compactification we are considering. In this paper
we will consider orbifolds which
break the supersymmetry to $N=1$. We will then resolve all the orbifold
singularities, which will give $M$-theory compactifications on smooth Joyce
8-manifolds of $Spin(7)$ holonomy \cite{J3}. As in \cite{hor,sen,Morb,sen1}, 
and by
analogy with string theory, we can assume that $M$-theory on an orbifold
has twisted sectors which are a necessary requirement for consistency
of the theory. We will further assume, again by analogy with string theory, 
that the massless fields associated with the resolution of orbifold
singularities constitute precisely these twisted sectors of $M$-theory.
It is possible that these twisted sectors will have a $p$-brane interpretion
\cite{Morb,sen,sen1} in future. 

A crucial point is the
following: because the coordinate labels of the $T^7$ compactified
heterotic string are a subset of the dual ${{T^4}/{Z_2}}{\times}T^4$ 
$M$-theory compactification, these further orbifolds, which break $N=8$ to
$N=1$, also define orbifolds of the heterotic string on $T^7$. To this
end, these further orbifolds must only act on these seven ``common''
coordinates, if the orbifolding is to commute with the original duality.
This was the case for the dual pairs conjectured in \cite{Ach}. Further,
because we resolve all singularities in the $M$-theory compactification,
it is natural to do this in the heterotic compactification. We thus
have the following picture, which summarises the construction of the dual
pairs in the following sections:

$M$-theory on {\Large $\frac{{S^1}{\times}{T^7}}{{Z_2}{\times}\Theta}$}
is dual to the heterotic string theory on {\Large $\frac{T^7}{\Theta}$.}
\nl
Here, $Z_2$ is the original $Z_2$ which acts as reflection on a $T^4$ factor,
giving $M$-theory on an orbifold limit of $K3{\times}T^4$, which is the
$N=8$ theory dual to the heterotic string on $T^7$. This $Z_2$ is the
only generator of the orbifold group which acts on the $S^1$ whose coordinate
label is {\it not} common to both compactifications. $\Theta$ denotes the
orbifold group which breaks supersymmetry from $N=8$ to $N=1$. The definition
of $\Theta$ on the $M$-theory coordinates then defines its action on the
$T^7$ coordinates of the heterotic theory. We will see in the next section
that the heterotic compactification also has $N=1$ supersymmetry on
general grounds.

In the next
section we consider a simple example in some detail. In section three we
show that our ansatz allows one to construct heterotic dual compactifications
for $M$-theory compactifications on all known, to date, Joyce 8-manifolds of
$Spin(7)$ holonomy \cite{J3}.

\section{Construction.}
The long wavelength dynamics of $M$-theory is effectively described by
eleven dimensional supergravity. This has a bosonic massless field
content consisting solely of the metric and three-form potential. 
Compactification
of the eleven dimensional theory on 8-manifolds of $Spin(7)$ holonomy
was considered in \cite{pap}. This gives a three-dimensional supergravity
theory with one supersymmetry in the vacuum. 
The non-trivial Betti numbers of a $Spin(7)$
Joyce manifold are $b_2$, $b_3$ and $b_4$; with $b_1=0$, $b_4$$={b_4^+} +
{b_4^-}$, where $b_4^+$ and $b_4^-$ are the dimensions of the self-dual and
anti-self-dual pieces of ${H^4}(X)$ respectively, where $X$ is the Joyce
manifold. The dimension of the
moduli space of such a Joyce manifold is $b_4^-$$+1$ \cite{J3}. Thus, 
compactification of the eleven dimensional theory on a Joyce $Spin(7)$
8-manifold leads to a three dimensional $N=1$ vacuum with 
$b_2$$+b_3$$+b_4^-$$+1$ massless scalar supermultiplets \cite{pap}.
In this paper we will apply our ansatz to $M$-theory compactified on
Joyce 8-manifolds of $Spin(7)$ holonomy. We will restrict our attention 
to the study of the simplest example of such a manifold \cite{J3}, and
present our full results in the next section.

In \cite{J3}, Joyce constructed many examples of $Spin(7)$ 8-manifolds as
blown up orbifolds of the eight torus, $T^8$. As far as we are aware, these
are, to date, the only known manifolds of this type. As in \cite{J3}, let
us denote the finite group, by which one is modding out, by $\Gamma$;ie 
in the notation of the introduction $\Gamma \equiv {Z_2}\times\Theta$.
It was
realised in \cite{J3} that there arise essentially five types of singularity
in the space ${T^8}/{\Gamma}$ which one has to blow up to construct a
Joyce 8-manifold of $Spin(7)$ holonomy\footnote{The definition of these
singularities will be irrelevant to this paper. The interested reader may
consult \cite{J3} for details.}. Each of these different blow ups
contribute different numbers of massless scalars to the $M$-theory
compactification. We will label these blow ups by {\it (i) - (v)}.
Of the five types, three have a unique resolution. They make the following
contribution to the Betti numbers of the manifold:
\begin{center}
\bq
Type {\it (i)}:\;adds\;1\;to\;b_{2},\;4\;to\;b_{3},\;
3\;to\;b_{4}^{+},\;3\;to\;b_{4}^{-}. \\
Type {\it (ii)}:\;adds\;1\;to\;b_{2},\;3\;to\;b_{4}^{+},\;3\;to\;b_{4}^{-}.\\
Type {\it (iii)}:\;adds\;1\;to\;b_{4}^{+}.
\eq
\end{center}
For each of the other two types there exist two topologically distinct 
resolutions of the singularity:
\bq
Type {\it (iv)}\;resolution(A):\;adds\;1\;to\;b_{2},\;2\;to\;b_{3},\; 
1\;to\;b_{4}^{+},\;1\;to\;b_{4}^{-}. \\
Type {\it (iv)}\;resolution(B):\;adds\;2\;to\;each\;of\;b_{3},\;
b_{4}^{+},\; b_{4}^{-}.
\eq
\bq
Type {\it (v)}\;resolution(A):\;adds\;1\;to\;each\;of\;b_{2},\;
b_{4}^{+},\; b_{4}^{-}. \\
Type {\it (v)}\;resolution(B):\;adds\;2\;to\;both\;b_{4}^{+},\;b_{4}^{-}.
\eq

Begin with $M$-theory on $T^8$. Orbifold
this theory by the $Z_2$ isometry denoted by $\alpha$, defined as follows:
\be
\alpha(x_1,x_2,...x_8) = (-x_1,-x_2,-x_3,-x_4,x_5,x_6,x_7,x_8)
\ee
where $(x_1,..x_8)$ are the coordinates of $T^8$.
Resolving the sixteen singularities associated with $\alpha$ gives $M$-theory
on $K3{\times}{T^4}$, which we expect to be equivalent \cite{wit} to
the heterotic string on $T^7$.

The crux of the ansatz \cite{Ach} which we explained in the 
introduction following Sen \cite{sen}, is that the
$T^7$ coordinates are labelled by a subset of $(x_1,...x_8)$. For definiteness
take the labels of $T^7$ as\footnote{According to our discussion in the
introduction, the $T^7$ coordinate labels must contain $x_5,x_6,x_7,x_8$ and
{\it any} three of $x_1,x_2,x_3,x_4$. We choose these to be $x_1,x_2,x_3$ for
ease of compatibility with \cite{J3}. However, we are of course free to
choose any three labels. This only means that the group $\Theta$ which
we further mod out by must be suitably modified so that it acts {\it solely} 
on the chosen seven labels.}
\be
(x_1,x_2,x_3,x_5,x_6,x_7,x_8)
\ee
Any further orbifolds of the $M$-theory geometry will then also be orbifolds
of the heterotic geometry.
In order to avoid confusion between the heterotic and $M$-theory geometries
later in the paper, we will re-label the $T^7$ coordinates of the heterotic
string as follows:
\be
(x_1,x_2,x_3,x_5,x_6,x_7,x_8) \equiv (y_1,y_2,y_3,y_4,y_5,y_6,y_7)
\ee
so that from now on $x_{i}$ will label coordinates of the $M$-theory
background and $y_{i}$ those of the heterotic background. We therefore
have $M$-theory on a $K3{\times}T^{4}$
background specified by $\alpha$ and $(x_{1},..x_{8})$ and the heterotic
string on a seven torus labelled by $(y_{1},...y_{7})$. When we speak of
$T^8$, it is implied that we are discussing the $M$-theory compactification;
similarly, when we speak of $T^7$, we are implicitly discussing the heterotic
compactification.
In what follows we will
take further orbifolds of these two $d=3$, $N=8$ theories (which will
give rise to $N=1$ vacua in three dimensions), resolve all singularities 
and consider the massless spectra. 

Consider then the following $\Gamma \equiv$ $({Z_2})^{4}$ $\equiv$ $(\alpha,
\Theta)$ orbifold 
of $T^8$ generated by the
following isometries:
\be
\alpha(x_1,...x_8) = (-x_1,-x_2,-x_3,-x_4,x_5,x_6,x_7,x_8)
\ee
\be
\beta(x_1,...x_8) = (x_1,x_2,x_3,x_4,-x_5,-x_6,-x_7,-x_8)
\ee
\be
\gamma(x_1,...x_8) = (c_1-x_1,c_2-x_2,x_3,x_4,c_5-x_5,c_6-x_6,x_7,x_8)
\ee
\be
\delta(x_1,...x_8) = (d_1-x_1,x_2,d_3-x_3,x_4,d_5-x_5,x_6,d_7-x_7,x_8)
\ee
where the $c_i$ and $d_i$ are constants which remain to be specified and
$\alpha$ is precisely the $Z_2$ element which defines the $K3{\times}T^{4}$
background dual to the heterotic string on $T^7$. According to our
ansatz then, $\Gamma$ also defines the action on the heterotic string 
toroidally compactified to
three dimensions. This is given by $\Theta \equiv$ $({Z_2})^{3}$ generated
by $(\beta,\gamma,\delta)$. Thus, the heterotic dual compactification
according to our ansatz, will be on a blown up orbifold of $T^7$, with
orbifold group defined by $\Theta$.

It can be checked \cite{J3}
that $\Gamma$ preserves the torsion free $Spin(7)$ structure
defineable on $T^8$. This orbifold therefore has discrete holonomy
contained in $Spin(7)$. By considering specific values for the constants
$c_i$ and $d_i$ Joyce \cite{J3} encountered the five singularity types
mentioned above. Blowing up all of these singularities in each case
leads to a smooth compact 8-manifold of $Spin(7)$ holonomy.

Let us consider the ``untwisted'' sector of $M$-theory on the orbifold
defined by equations (11)-(14). To compute the massless spectrum, it
suffices to calculate the Betti numbers of ${T^8}/{\Gamma}$. These are
computed to be:
$b_{1}=b_{2}=b_{3}=0$ and
$b_{4}^{+}=b_{4}^{-}=7$. This is {\it always} the case for {\it any} choice
of the constants $c_i$ and $d_i$. This implies that $M$-theory on the orbifold
defined by equations (11)-(14) has an untwisted sector consisting of $N=1$
three-dimensional supergravity coupled to eight scalar multiplets. We will
see shortly the interpretation of this fact in terms of the dual heterotic
theory.

The first example considered in \cite{J3} is the following. Set
\be
(c_1,c_2,c_5,c_6) = (1/2,1/2,1/2,1/2)
\ee
and
\be
(d_1,d_3,d_5,d_7) = (0,1/2,1/2,1/2)
\ee

In this example, it may be checked \cite{J3}
that the singular set
of ${T^8}/{\Gamma}$ contains four singularities of Type {\it (i)}, eight
of Type {\it (ii)} and 64 of Type {\it (iii)}. Some simple arithmetic 
shows that the non-zero Betti numbers of the smooth $Spin(7)$ manifold
are
\be
b_{2} = 12, b_{3} = 16, b_{4}^{-} = 43, b_{4}^{+} = 107
\ee

We can, by analogy with string theory, consider the massless fields coming
from the blowing up modes of the $Spin(7)$ manifold, as constituting twisted
sector states of the $M$-theory background; ie in addition to the 
``universal'' eight scalar multiplets from the untwisted sector, we have
64 from the twisted sector.
This means that $M$-theory compactified on such a manifold gives a 
three-dimensional $N=1$ theory with $72$ scalar multiplets.

Let us now see what this implies for the heterotic theory. According to
our ansatz, the finite group $\Theta \equiv$ ${Z_2}^{3}$ will act on the
heterotic $T^7$ coordinates as follows:

\be
\beta(y_1...y_7) = (y_1,y_2,y_3,-y_4,-y_5,-y_6,-y_7)
\ee
\be
\gamma(y_1,...y_7) = (c_1-y_1,c_2-y_2,y_3,c_5-y_4,c_6-y_5,y_6,y_7)
\ee
\be
\delta(y_1,...y_8) = (d_1-y_1,y_2,d_3-y_3,d_5-y_4,y_5,d_7-y_6,y_7)
\ee
It is not too difficult to show \cite{J1,J2} that, {\it in general} 
the resolution
of the singularities in ${T^7}/{\Theta}$ will {\it always} give rise to a
Joyce 7-manifold of $G_2$ holonomy, for {\it any} choice of the constants, 
$c_i$ and $d_i$. This will give rise to a heterotic
background in three dimensions which always has $N=1$ supersymmetry. This is
consistent with the fact that the dual $M$-theory compactification also has
one supersymmetry in three dimensions. This fact is a highly non-trivial
statement because the isometry group defined by equations (18)-(20) is
essentially the {\it only} ${(Z_2)}^3$ group which gives rise to the correct 
holonomy (ie $G_2$) for the heterotic compactification space, for a specific
choice of $G_2$ structure \cite{J3}.

Let us consider, as we did above 
for the $M$-theory compactification, the ``orbifold interpretation''. In the
untwisted sector of the heterotic string on the orbifold defined by equations
(18)-(20), one finds for the massless modes three dimensional supergravity
coupled to eight scalar multiplets. This is precisely in accord with what
we found for the dual $M$-theory result above\footnote{The heterotic theory
also has 16 vector multiplets which are dual to scalar multiplets in three
dimensions. However, the $M$-theory duals of these vectors have in all 
examples to date arisen from the twisted sector of the $M$-theory
compactification. So, it suffices to consider the ``matter'' sector.}.
However, different choices of the
constants $c_i$ and $d_i$ will lead to different Betti numbers for the
$M$-theory and heterotic backgrounds respectively and it would be truly
remarkable if the massless spectra in the two theories are the same. We will
demonstrate that this is indeed the case for all the $Spin(7)$ 8-manifolds 
constructed by Joyce \cite{J3}. Let
us consider our example. 

It can be checked \cite{J2} that for the specific choice of constants
$c_i$ and $d_i$ that we are considering that the resolution of orbifold
singularities on the heterotic side of the duality map gives the heterotic
string on a Joyce $G_2$ manifold with Betti numbers $b_2$=$12$ and $b_3$=
$43$. In order to specify the heterotic background we must specify
the expectation value of the gauge fields of the heterotic string
on the Joyce manifold. In this paper, we will limit ourselves to
an abelian embedding of the spin connection in the gauge connection, such
that the heterotic gauge group, $SO(32)$ or $E_{8}{\times}E_{8}$ is
broken to its maximal abelian subgroup, as in \cite{pap}\footnote
{This choice of embedding is possible because the Joyce manifolds of
$G_2$ holonomy which we consider in this paper contain $K3$ submanifolds
\cite{J2} (or see \cite{Ach}). If we restrict the gauge fields to take
expectation values on a $K3$ submanifold, then it is known that abelian
embeddings are consistent \cite{green}. It is of course possible that
abelian embeddings are consistent for Joyce manifolds in much more
generality.}. With this choice of embedding, the 
massless spectrum of the heterotic compactification is an 
$N=1$ theory in three dimensions with
$72$ scalar multiplets, precisely what is expected by duality!

\section{A Heterotic Dual for All $M$-theory Compactifications on Joyce
$Spin(7)$ Manifolds.}

The $Spin(7)$ manifold constructed in the previous section is the simplest
example of \cite{J3}. This was because the singular set of ${T^8}/{\Gamma}$
contained elements of Type {\it (i)-(iii)}, for each of which there exists
a unique resolution. All the other examples in \cite{J3} contain, in addition, 
singular elements of Type {\it (iv)} and {\it (v)}, for each of which there 
exist two topologically distinct choices of resolution. It follows that these
examples will consist of finite families of Joyce $Spin(7)$ manifolds,
labelled by an integer parametrising the choice of resolution made for
each singularity for which a choice exists. As mentioned, the precise
details of the construction of the manifolds is not of immediate interest to
the present paper. Therefore, we will now present our results, which are
documented in the two tables below below.
\nl
\nl
\begin{tabular}{|l|l|l|l|l|l|l|l|l|l|}   \hline\hline

Example&{\bf $c_i$}&{\bf $d_i$}&\multicolumn{4}{c|}{$Spin(7)$}
&\multicolumn{2}{c|}{$G_2$}&Scalars\\
\hline
 &$(c_{1},..c_{4})$&$(d_{1},..d_{4})$&$b_{2}$&$b_{3}$&$b_{4}^{+}$&$b_{4}^{-}$
&$b_{2}$&$b_{3}$&n\\
\hline
1.&$((1/2)^{4})$&$(0,(1/2)^{3})$&12&16&107&43&12&43&72\\
\hline
2.&$(1/2,0,1/2,0)$&$(0,(1/2)^{3})$&10+j&16&109-j&45-j&12&43&72 \\
\hline
3.&$((1/2)^{3},0)$&$(0,1/2,1/2,0)$&8+k&16&111-k&47-k&8+l&47-l&72 \\
\hline
4.&$(1/2,0,1/2,0)$&$(0,1/2,1/2,0)$&6+m&16&113-m&49-m&8+l&47-l&72 \\
\hline
\end{tabular}
\nl
\nl
Table1: Examples of dual $N=1$ 
$M$-theory and heterotic compactifications to three
dimensions.

In Table 1, the columns labelled $c_i$ and $d_i$ specify the orbifold
isometry groups ${\Gamma}$ for $M$-theory and ${\Theta}$ for the heterotic
string, according to equations (11)-(14) and (18)-(20) respectively. 
The column labelled $Spin(7)$ gives the Betti numbers of the smooth
Joyce 8-manifold of $Spin(7)$ holonomy on which $M$-theory is
compactified. The column labelled $G_2$ gives the Betti numbers of the
smooth Joyce 7-manifold of $G_2$ holonomy on which the heterotic theory
is compactified. The integers $j,k,l,m$ range from $0$ to $4,8,8,12$ 
respectively. The last column gives the number, $n$ of massless $N=1$
scalar multiplets in three dimensions. For the $M$-theory compactification
$n=$$b_2$$+b_3$$+b_4^-$$+1$, whereas for the heterotic compactification,
$n=$$b_2$$+b_3$$+17$. It is remarkable that the number of
such multiplets agrees for {\it both} theories.

In \cite{J3} Joyce goes on to consider a further $Z_2$ orbifold of some
of the above $Spin(7)$ manifolds and produces further examples of such
manifolds. This further $Z_2$ orbifold is generated by the following
isometry:
\be
\epsilon(x_1,...x_8) = (c_1+x_1,c_2+x_2,x_3,1/2+x_4,
c_5+x_5,c_6+x_6,x_7,1/2+x_8)
\ee
The constants $c_i$ appearing in this equation are the {\it same} constants
which appeared in equation (13). Thus, if we consider our first example
again, which had $c_i$ = $({(1/2)^4})$, then we can take a further orbifold of
this manifold, with isometry generated by $\epsilon$ and resolve all 
singularities. Because the element $\epsilon$ acts freely on all coordinates,
the torsion free $Spin(7)$ structure is left invariant. Thus the manifolds
produced this way will also be Joyce $Spin(7)$ manifolds. Because of our
ansatz, the isometry $\epsilon$ will also define a $Z_2$ orbifold of the
heterotic string on the Joyce $G_2$ manifolds in Table 1 above. It is easily
seen along similar lines that the action of $\epsilon$ preserves the $G_2$
structure of the heterotic geometry as well. We document the results
of two additional examples considered in \cite{J3} in Table 2 below.
\nl
\nl

\begin{tabular}{|l|l|l|l|l|l|l|l|l|} \hline \hline
\multicolumn{2}{c|}{{\bf Example}}&\multicolumn{4}{c|}{$Spin(7)$}
&\multicolumn{2}{c|}{$G_2$}&Scalars\\
\hline
 &Ex.$+\epsilon$&$b_2$&$b_3$&$b_{4}^{+}$&$b_{4}^{-}$&$b_2$&$b_3$&n\\
 \hline
5&$1+\epsilon$&9&4&98&34&6&25&48\\
\hline
6&$4+\epsilon$&4+n&4&103-n&39-n&2+l&29-l&48\\
\hline
\end{tabular}
\nl
\nl
Table 2: Further examples of dual $M$-theory/heterotic compactifications.

In Table 2 the second column denotes which manifold in Table 1 is being
further orbifolded by $\epsilon$. In this table, the integers $n,l$ range
from $0$ to $10,4$ respectively. Again we see that both theories have the
same massless spectra. An interesting illustration of the remarkable nature
of these results comes from example 6, Table 2. In this example the orbifold
action on the heterotic geometry given in our ansatz was such that the
resulting smooth Joyce manifold of $G_2$ holonomy was not one which has
appeared in \cite{J1,J2}. Our ansatz nevertheless
succeeded in not only producing a new family of Joyce manifolds of $G_2$
holonomy, but precisely a family which gives the correct massless heterotic
spectrum as required by duality.

Finally, there was one more family of Joyce $Spin(7)$ manifolds constructed
in \cite{J3}. The orbifold group in this example was the $({Z_2}^{5})$ group
generated by $(\alpha,\beta,\gamma,\delta,\epsilon)$, as defined above.
$c_i$ and $d_i$ were chosen for this example to be:
\bq
(c_1,..c_4) = (0,1/2,0,1/2) \\
(d_1,..d_4) = (1/2,0,0,1/2)
\eq
This example leads to a compactification of $M$-theory on a Joyce 8-manifold
of $Spin(7)$ holonomy with Betti numbers given by:

$b_{2}=$ $8+j$; $b_{3}=8$; $b_{4}^{+}=103-j$; $b_{4}^{-}=39-j$, for $j=0,..4$.

This example leads to a three-dimensional theory with $56$ scalar multiplets.
Applying our ansatz to construct the heterotic dual, we find the heterotic
string compactified on a Joyce 7-manifold of $G_2$ holonomy, with betti
numbers $b_{2}=$ $4+l$ and $b_{3}=$ $35-l$, for $l=0,..8$. 
This also leads to an $N=1$
theory in three dimensions with $56$ scalar multiplets.

We have thus demonstrated the consistency of our ansatz, and constructed
heterotic dual compactifications for all known $M$-theory compactifications
on Joyce 8-manifolds of $Spin(7)$ holonomy.

\section{Discussion and Comments.}

We have presented strong evidence for the existence of heterotic duals
for all known $M$-theory compactifications on Joyce manifolds of $Spin(7)$
holonomy. The heterotic duals are compactifications on Joyce manifolds of
$G_2$ holonomy. All Joyce manifolds constructed to date \cite{J1,J2,J3} are
based on the blown up orbifold construction. This fact was utilised in
\cite{shat}, where string compactifications on these spaces were considered
as orbifold conformal field theories. This of course applies to the heterotic
compactifications considered here. Most of the Joyce orbifolds we have
considered here possess orbifold singularities which admit more than one
resolution. This leads to the families of manifolds that we have discussed.

It was found in \cite{shat} that string theories compactified on different
Joyce manifolds from the {\it same} family give equivalent conformal field
theories up to deformations in the moduli space. 
Specifically, for string compactification on Joyce $G_2$ orbifolds,
equivalent conformal field theories have the same ${b_2}+{b_3}$. This means
that all the heterotic compactifications in Table 1 are equivalent up
to moduli deformation. 
This in turn implies that the four families of $M$-theory
compactifications are also equivalent up to deformation. An argument which
supports
this statement is the following: Compactify the theories in Table 1 on an
$S^1$. These should then be equivalent \cite{wit} to $Spin(7)$ 
compactifications of Type IIa string theory. However, from \cite{shat}, one
learns that all these string compactifications are equivalent as
conformal field theories, up to changes in the moduli.
Then take the strong coupling limit of the Type IIa theory, and we will
recover three dimensional Lorentz invariance \cite{wit}. In this limit
the Type IIa compactification is described by the weakly coupled $M$-theory
compactification in Table 1. This is just a particular limit in the moduli
space of the Type IIa theory, in which the moduli of the Joyce manifold
are not varied. Hence, the $M$-theory compactifications of Table 1 
should also be equivalent\footnote{However, it should be noted that the work
of \cite{shat} was carried out in the formalism of string perturbation
theory and it is not clear to what extent the results of \cite{shat}
may be extrapolated to strong coupling. Our results do seem to indicate that 
the results of \cite{shat} do hold in this limit.}.
The same reasoning of course applies to
all the compactifications in Table 2. In fact, if the above were not true,
then the duality proposed in this paper would be less concrete. For example,
consider consider Table 1, row 4. Here we propose $13$ (labelled by $k$)
compactifications of $M$-theory dual to $9$ (labelled by $l$) 
compactifications of heterotic string theory. If it were not for the preceding
comments, one would be left wondering, how is $k$ related to $l$? However,
given our comments, we do not have to worry about such a question as these
parameters do not have any bearing on the moduli space of these
theories. This, presumably, also follows from the number of scalar fields
present in three dimensions, together with information concerning the 
structure of the moduli spaces of Joyce manifolds. Unfortunately such
knowledge is not yet available to check this.

Much progress has recently been made in discussing $M$-theory on orbifolds
in dimensions where anomaly cancellation arguments 
are useful \cite{hor,Morb,sen,sen1}. In particular, in \cite{hor,sen,sen1} 
and the first reference of \cite{Morb},
$p$-branes played a crucial role in determining the full twisted sector
spectrum. An important open question is do $p$-branes in $M$-theory play
an analogous role here. It is certainly natural to speculate that they do.

Finally, we wish to comment on the relationship between this work and possible
physics in twelve dimensions. Recently \cite{Y,F}(see also \cite{k} 
and references therein), 
the existence of a mysterious theory in twelve dimensions
has been speculated upon. It is not yet clear what the relationships between
these various ideas is, but in \cite{Y,F} it was conjectured that
compactification of a twelve dimensional theory on a circle is equivalent
to $M$-theory in eleven dimensions. Using this fact, it was realised in
\cite{F} that compactification of the twelve dimensional theory on
a Joyce manifold of $Spin(7)$ holonomy apparently gives a theory with
no supersymmetry in four dimensions. This is because further $S^1$
compactification gives the compactifications considered in this paper.
It was pointed out in \cite{F} that this could be an explicit realisation
of the ideas of Witten $\cite{cos}$, which may solve the cosmological
constant problem together with the problem of bose-fermi mass degeneracy, in
a supersymmetric context. If this is indeed the case, then the 
compactifications considered in this paper deserve yet further study. 
One possible
avenue for this is to use Type I - heterotic duality \cite{het} and
consider the resulting Type I compactifications on the same Joyce
$G_2$ manifolds that we have considered here. 
One advantage of this approach is that one can study
non-perturbative Type I physics in these backgrounds along
the lines of \cite{D}. We hope to report on these issues in the near
future.

{\bf Acknowledgements.}

The author would like to thank M.Abouzeid,
J.Gauntlett, C.M.Hull, J.Polchinski and
A.Sen for useful discussions; and PPARC, by whom this work is supported.

\end{document}